\documentclass[preprint,prb,twoside,amsmath,aps]{revtex4}

\usepackage{graphicx}

\begin{document}

\title{THz frequency- and wavevector-dependent conductivity of
low-density drifting electron gas in GaN. Monte Carlo calculations.}

\author{G. I. Syngayivska}
\affiliation{Department of Theoretical Physics, Institute of Semiconductor Physics of NAS of Ukraine,  03028 Kyiv,
Ukraine}

\author{V. V. Korotyeyev\footnote{koroteev@ukr.net. This article may be downloaded for personal use only. Any other use requires prior permission of the author and AIP Publishing. This article appeared in (J. Appl. Phys. 125, 135704 (2019)) and may be found at (https://doi.org/10.1063/1.5082016)}}
\affiliation{Department of Theoretical Physics, Institute of Semiconductor Physics of NAS of Ukraine,  03028 Kyiv,
Ukraine}

\author{V. A. Kochelap}
\affiliation{Department of Theoretical Physics, Institute of Semiconductor Physics of NAS of Ukraine,  03028 Kyiv,
Ukraine}

\author{L. Varani}
\affiliation{Institute of Electronics and Systems, UMR CNRS 5214, University of Montpellier,  France}

\begin{abstract}
We report the results of Monte Carlo simulation of electron dynamics in
stationary and space- and time-dependent electric fields in compensated GaN samples.
We have determined the frequency and wavevector dependencies of the dynamic conductivity,
$\sigma_{\omega,q}$. We have found that the spatially dependent dynamic conductivity
of the drifting electrons can be negative under stationary electric fields of
moderate amplitudes, $2..5$ kV/cm. This effect is realized in a set of frequency windows.
The low-frequency window with negative dynamic conductivity is due to the
Cherenkov mechanism. For this case the time-dependent field induces a {\it traveling wave} of the electron concentration
in real space and a {\it standing wave} in the energy/momentum space.
The higher frequency windows of negative dynamic conductivity are associated with
the optical phonon transient time resonances. For this case the time-dependent field
is accompanied by oscillations of the electron distribution in the form of
the {\it traveling} waves in both the real space and the energy/momentum space.
We discuss the optimal conditions for the observation of these
effects. We suggest that the studied negative dynamic conductivity can be used
to amplify electromagnetic waves at the expense of energy of the stationary
field and current.
\end{abstract}

\maketitle

\section{Introduction}\label{Introduction}

In polar semiconductor materials and heterostructures,
such as III-V compounds, group-III nitrides, ZnO/MgO and others, at low lattice temperature
the optical phonon emission is the dominant scattering mechanism for {\it hot electrons},
which considerably suppresses their mobility.
Meanwhile the electrons can have a high {\it low-field mobility}.  Indeed,
at low temperatures, when $e^{-\hbar \omega_{op}/k_B T_{0}} \ll 1$
($\omega_{op}$,  $k_B$ and $T_{0}$ are the optical phonon frequency,
the Boltzmann constant and the temperature, respectively)
the absorption/emission of optical phonons by the equilibrium electrons
is practically absent and the electron mobility is limited only by weak
quasi-elastic scattering by impurities and acoustic phonons.
Under these conditions the dynamics of an electron subjected
to a steady-state high electric field $F_0$ is the following.
The electron is almost ballistically accelerated by the field
until reaching the optical phonon energy, $\hbar\omega_{op}$. Then, an optical
phonon emission occurs so that the electron looses practically
all its energy and stops, then this process is repeated again.
This electron dynamics gives rise to temporal and spatial modulation
of the electron momentum, ${\bf p}$, velocity, ${\bf v}$, and concentration, $n_e$,
with characteristic time period,
$\tau_F = p_{op}/e F_0$, and space period, $l_F = e F_0
 \tau^2_F/2\,m^* \equiv \hbar \omega_{op}/e F_0$, where
 $p_{op} =\sqrt{2 m^* \hbar \omega_{op}}$, $e$ is the elementary charge
 and $m^*$ is the electron effective mass.
This is essentially a single-electron physical picture, which is valid at
low or modest electron concentrations, when {\em e-e} collisions
do not destroy the cyclic motion. Note that such a
{\it cyclic electron dynamics} in real and momentum/energy spaces
due to strong scattering by optical phonons
was predicted many decades ago by Shockley.~\cite{Shockley}

Experimental evidences of the cyclic dynamics
in real space were found by analyzing low temperature I-V
characteristics of short diodes made from different polar materials:
InSb,\cite{InSb} InGaAs,\cite{InGaAs}  GaAs,\cite{GaAs} and
InP \cite{InP}. At low temperatures tens of cycles were identified.
For electrically biased short InN and GaN diodes,
the formation of stationary one-dimensional gratings of electron concentration
and velocity was predicted for nitrogen temperature in Refs.~[\cite{Reggiani-1,Gonzalez-1}].

In the {\it frequency domain}, the cyclic  electron dynamics gives
rise to a resonance phenomenon at the transit-time frequency
$\omega_F = 2 \pi /\tau_F$,  frequently
called optical phonon transit time  (OPTT) resonance.
Among a number of interesting effects induced by the OPTT resonance (see
Refs.~[\onlinecite{Andronov}], [\onlinecite{Reggiani-review}]) the most
interesting is the appearance of a negative high-frequency (HF) conductivity, $\sigma (\omega)$,
of electrons at the frequencies, $\omega \sim \omega_F$,
which leads to the possibility of amplification and generation
of electromagnetic waves in the sub-THz and THz frequency regions.
The OPTT resonance generation was studied theoretically
in details for bulk materials~\cite{Andronov,Reggiani-review} and
low-dimensional heterostructures~\cite{2DEG-OPTTR,2DEG-OPTTR-2,2DEG-OPTTR-3,streaming-CNT}.
This type of high-frequency generation was observed experimentally
in InP samples for the frequency range $50$ to $300$ GHz.\cite{Vorobiev}

The cyclic electron dynamics also gives rise to a complex motion in the {\it
phase space} associated with time-periodic oscillations (waves)
of the electron concentration/charge in real space and synchronized
electron redistribution in momentum space. This results in a significant
 (resonant) spatio-temporal dispersion of the electron response to nonuniform
 electromagnetic waves with (angular) frequency, $\omega$ and wavevector $\bf  q$:
 $\sigma(\omega, \bf q)$. As shown in Ref.~\onlinecite{Korotyeyev},
the oscillations in the phase space can be realized as self-supporting
and weakly damped excitations of the drifting electron gas.
The excitations are quite different from the well known plasmons. Indeed,
their frequency-wavevector relations are presented by an infinite number of
continuous branches, $\omega^k (q)$, with $q$ being the wave vector
of the excitations and $k=0, \pm 1, \pm 2...$. The damping of these
oscillations is weak or even absent, when the frequency and/or
the wavevector are multiples of $\omega_F$ and/or
$q_F = 2 \pi /l_F$, respectively, i.e., under conditions of time- and/or space
resonances.

This novel type of spatio-temporal resonant phenomena was studied
analytically in~\cite{Korotyeyev} by using the approximation of
infinitely fast emission of optical phonons by the electrons
with energy exceeding $\hbar\omega_{op}$. In fact, a finite rate
of the electron relaxation on the optical phonons is critically
important. Indeed, this relaxation can limit the temperature
interval and the electric field range, where these resonances may be observed
and practically exploited.

In this paper, we present a numerical study of
the spatio-temporal dispersion of the HF conductivity $\sigma(\omega, q)$
under the OPTT resonance effect. The calculations were carried out in the framework
of the Monte Carlo method taking into account all actual relaxation processes.
As a result, we found and investigated wave-like excitations of
the electron gas and confirmed the existence of pronounced spatio-temporal resonances
in $\sigma(\omega, {\bf q})$ at $\omega \approx \omega_F$ and
$q \approx 2 \pi/l_F$ in perfect bulk GaN crystals subjected to
an  electric field of  moderate strength. Finally, we determined the
$\{\omega, q\}$-regions, where the real part of the HF conductivity
is negative, the drifting electron gas is unstable and an external
electromagnetic wave with corresponding $\omega$ and $q$ can be amplified
at the expense of the stationary field and current.

\section{Transport model}

The analysis of semiconductor materials with strong electron-optical phonon interaction
has showed that the group-III nitrides are among the most promising materials for the study, observation and application of the OPTT resonance phenomena~\cite{2DEG-OPTTR-2,2DEG-OPTTR-3}.
In this paper we consider a bulk-like  GaN sample with cubic lattice structure and given concentration of ionized impurities, $N_{i}$.
We assume that the sample is compensated to exclude quenching effect on
the OPTTR by electron-electron scattering, i.e. $n_{e}<N_{i}$ where  $n_{e}$ is the
electron concentration. At electric fields of moderate strength,
all electrons remain in the $\Gamma$ valley and can be characterized
by a parabolic dispersion law with effective mass $m^*=0.2\,m_0$,
where $m_0$ is the free electron mass.
The stationary, $F_0$, and alternating, $\tilde{F}$, electric fields are assumed
to be parallel  and both directed along  the $OZ$-axis.
The alternating field is assumed to be in the form of a wave propagating along the $OZ$-axis:
\begin{equation} \label{AF}
 \tilde{F}(z,t) = F_{\omega,q} \,\cos(q z-\omega t)\,.
\end{equation}
To find the small-signal response, the alternating field should be considered
as a small one: $|F_{\omega,q}| \ll | {F}_0|$.

To calculate the electron transport characteristics including the electron
distribution function, the current-voltage characteristics and the electron response, $\sigma(\omega, \bf q)$,
to the alternating field (\ref{AF}), we exploit the {\it single-particle Monte Carlo}
procedure~\cite{Boardman, Reggiani}, which is extensively used to solve a wide variety of problems involving transport at a kinetic level.
To simulate the electrons dynamics we use, as usual,  the Newton equation with
the force $-e[F_0+\tilde{F} (z,t)]$ to describe the free flight of the electron between two subsequent scatterings and
take into account three main scattering mechanisms: interactions with
ionized impurities, acoustic phonons and polar optical phonons.
For the ionized impurity scattering, we exploit the mixed scattering model
unifying the Brooks-Herring and Conwell-Weisskopf models. The latter approach
is more appropriate for the analysis of compensated materials~\cite{Reggiani},
on which our analysis is focalised.
The Monte Carlo simulation of electron transport in stationary fields
is a standard procedure whose application to GaN material can be found
elsewhere~\cite{Mc3,Mc4}.

In paper~\cite{Zimmerman}, the single particle Monte Carlo
algorithm was applied to the calculation of the electron response
to a time-periodic perturbation. We extended this approach to
the electron system subjected to both uniform stationary and time-
and space-dependent electric fields.
The details of the calculation algorithm, its accuracy, stability and convergence are discussed in the Appendix.

As an example, in Fig.\ref{fig1} (a) we present a 3D-plot of the alternating current,
$\tilde j(z,t)=J_{z}(z,t)-J_{z,0}$ within a single time and spatial periods of the alternating electric signal (see Eqs. (\ref{Current}) and (\ref{Current_dc}) in Appendix). These results are obtained for a stationary field
$F_0 = 3\,kV/cm$ and an alternating field with parameters: ${F}_{\omega,q} = 0.3$ kV/cm,
$\omega = 0.2$ THz, $q= 10^{5}$ cm$^{-1}$. The impurity concentration,
the electron concentration and the ambient temperature are $N_i = 10^{16}$ cm$^{-3}$,
$n_e = 10^{15}$ cm$^{-3}$ and $T_{0}=30$ K, respectively.
\begin{figure}[h]
\includegraphics[width=0.47\textwidth]{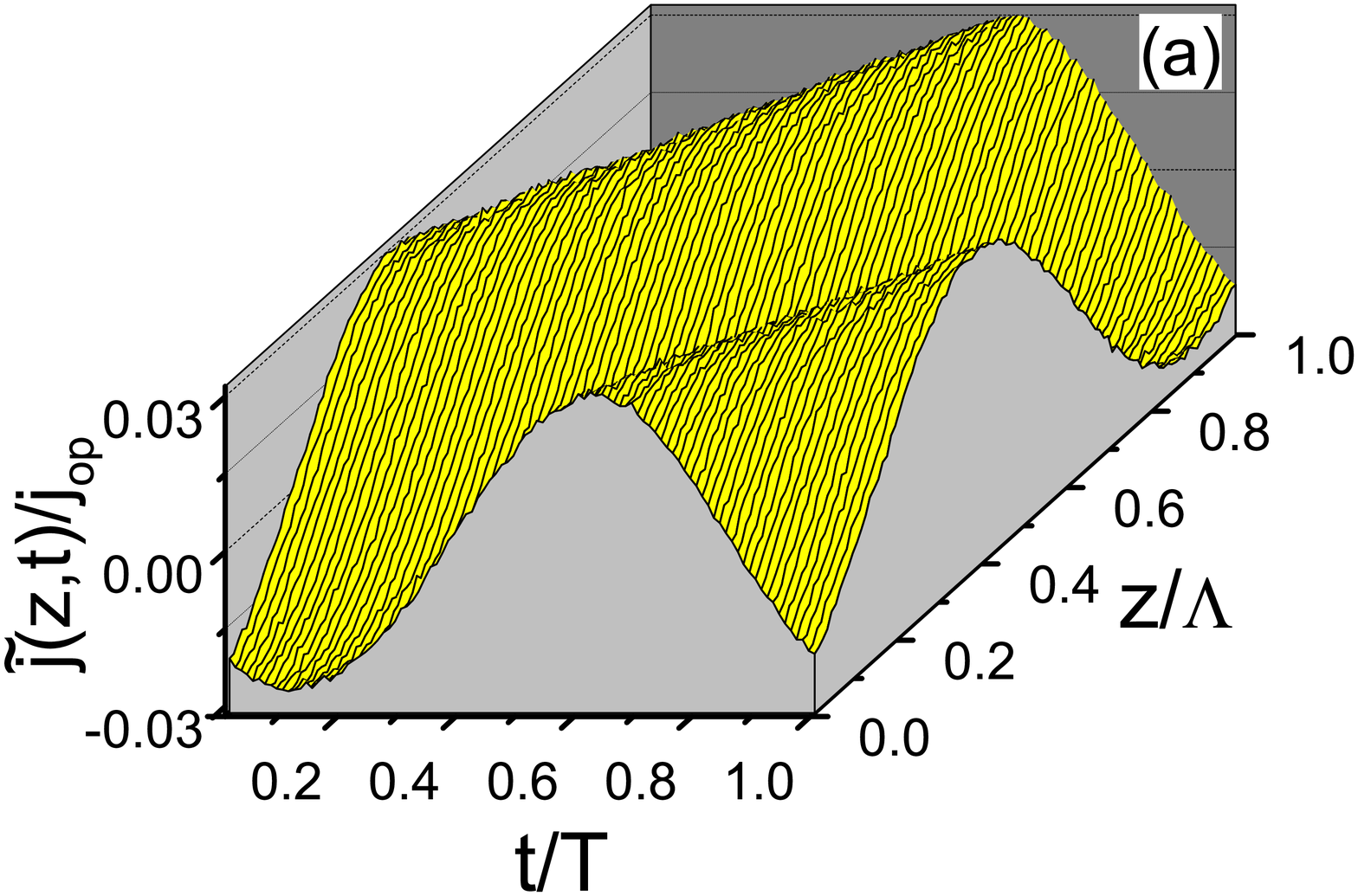}
\includegraphics[width=0.45\textwidth]{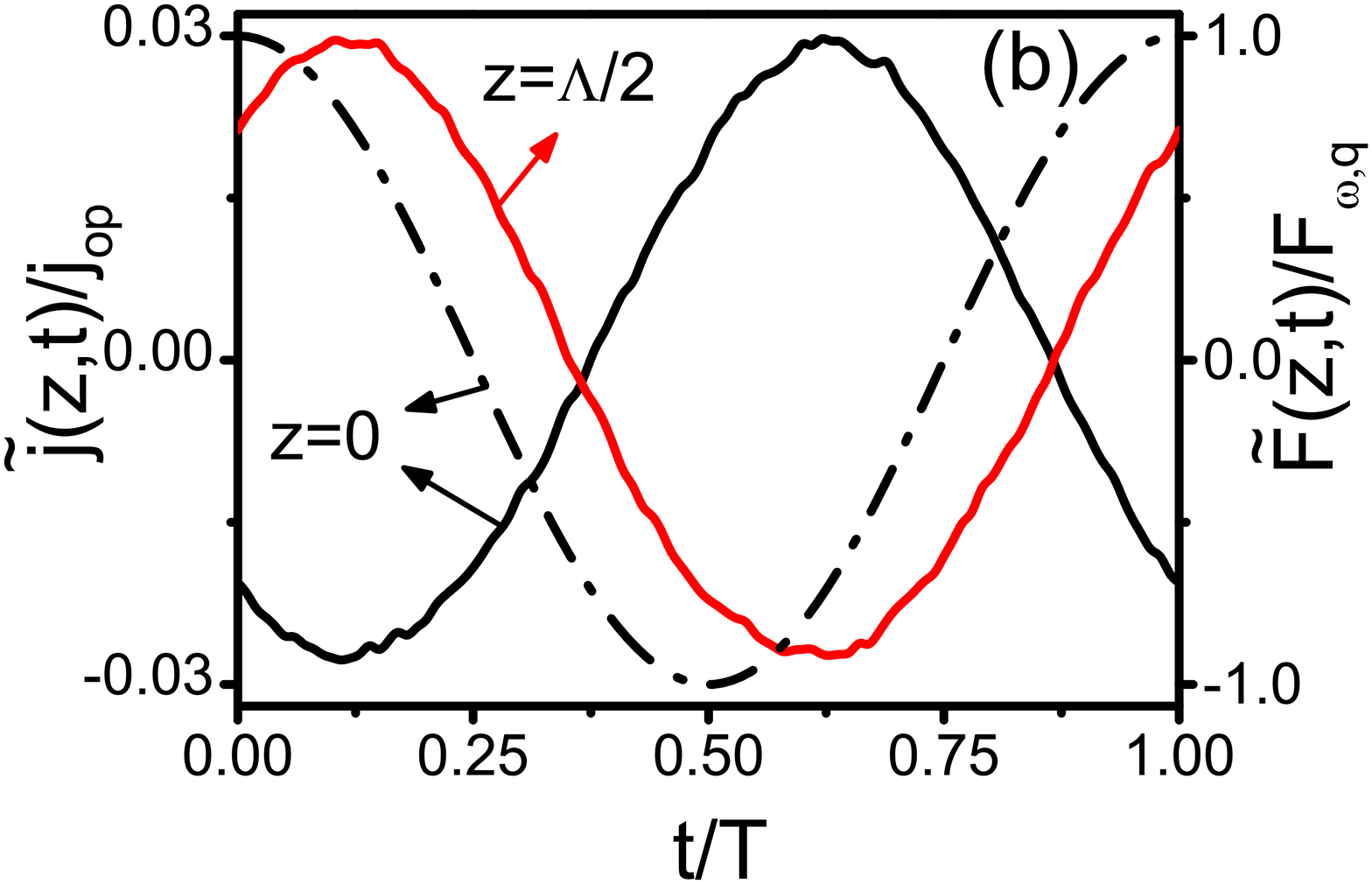}
\includegraphics[width=0.45\textwidth]{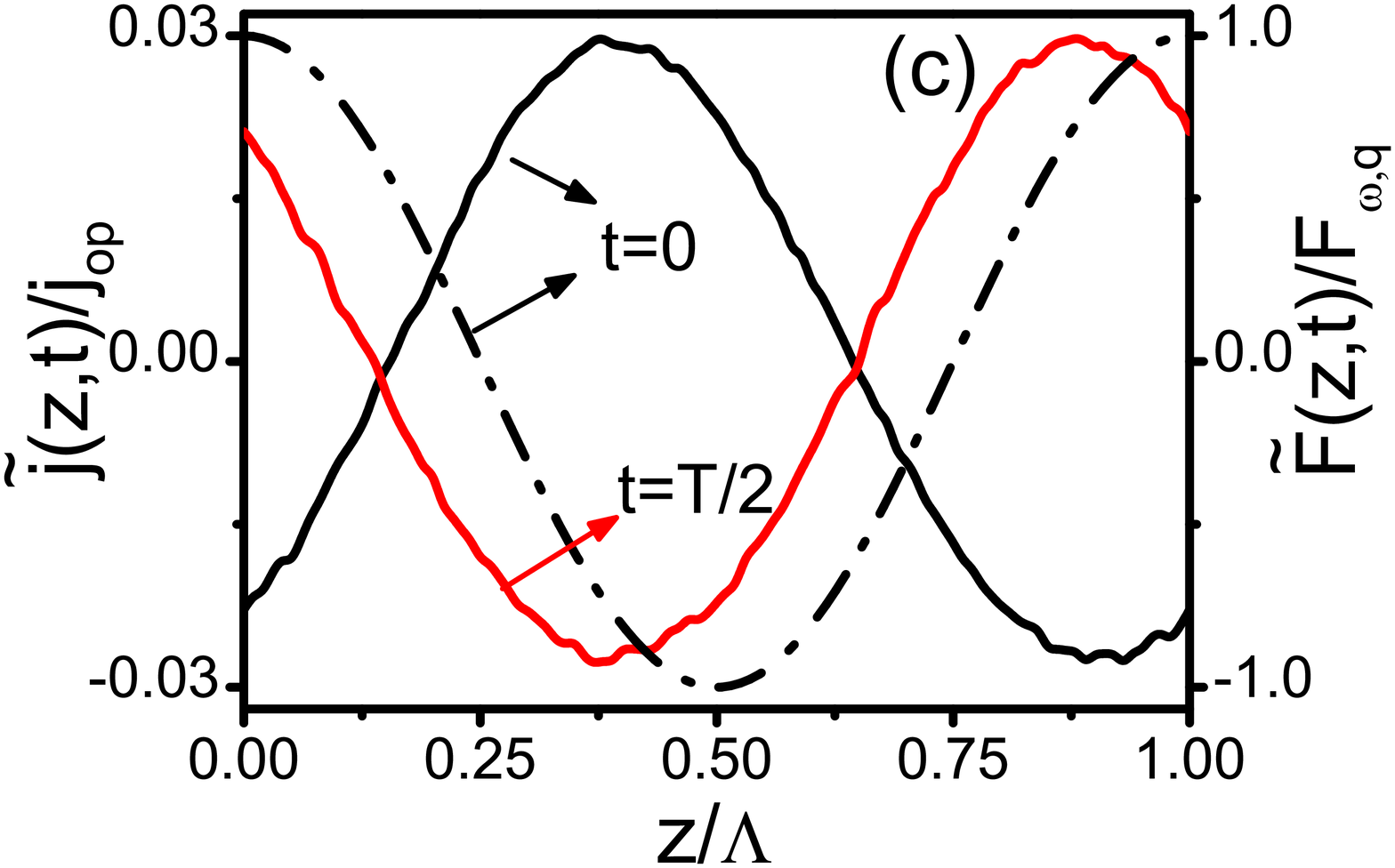}
\vskip 0.1 cm
\caption{(a): Alternating current ${\tilde j} (z,t)$ normalized to the characteristic current, $j_{op}=en_{e}p_{op}/m^{*}$.
(b): Time dependence of $\tilde j$ at a given $z$.
(c): Spatial dependence of $\tilde j$ at a given $t$.
The dash-dotted lines in (b) and (c) show the alternating electric field
for comparison. The simulation parameters are $\omega = 0.2$~THz, $q=10^5$~cm$^{-1}$, $T_{0}=30$~K, $F_{0}=3$~kV/cm, $F_{\omega,q}=0.1\times F_{0}$. }
\label{fig1}
\end{figure}

We remark that the alternating current $\tilde j(z,t)$ exhibits a nearly
plane-wave behavior. Figures~\ref{fig1} (b) and (c) allow to compare the spatial and temporal
dependencies of the alternating current with those of the wave field
${\tilde F} (z,t)$. From these figures one can conclude that
between the alternating current and the alternating field there is
a phase-shift $\Delta \varphi_{\omega,q}$.

Below we present the obtained results in terms of the {\it complex HF conductivity}.
Note that since $\sigma_{\omega,q}$ is the linear response to the field in the
form of Eq.~(\ref{AF}), we will use the following properties:
\begin{eqnarray}
Re[\sigma_{\omega,q}] =Re[\sigma_{-\omega, -q}],\,\,Re[\sigma_{\omega,-q}] = Re[\sigma_{-\omega, q}],\nonumber\\
Im[\sigma_{\omega,q}]=-Im[\sigma_{-\omega,-q}],\,\,Im[\sigma_{\omega,-q}]=-Im[\sigma_{-\omega,q}].\nonumber
\end{eqnarray}
Due to these relationships, we will present the result only for $q > 0$ while $\omega$
will take both positive and negative values.

\section{Frequency and wavevector dispersions of the HF conductivity}

The obtained HF conductivity, $\sigma_{\omega,q}$, is dependent on
the frequency, $\omega$, and the wavevector, $q$, i.e. both temporal and spatial
dispersions of the HF conductivity are important.

\begin{figure}[h]
\includegraphics[width=0.45\textwidth]{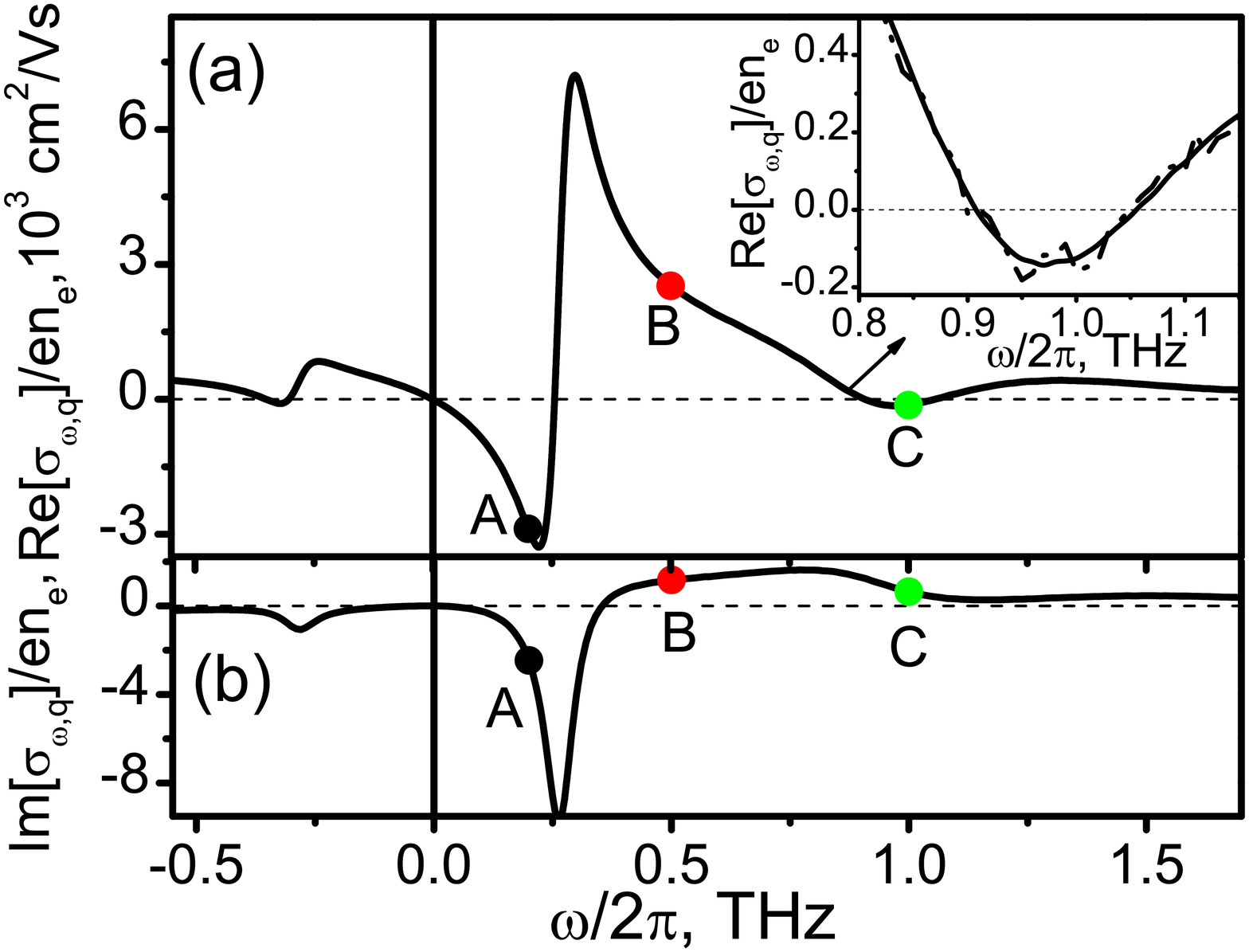}
\includegraphics[width=0.45\textwidth]{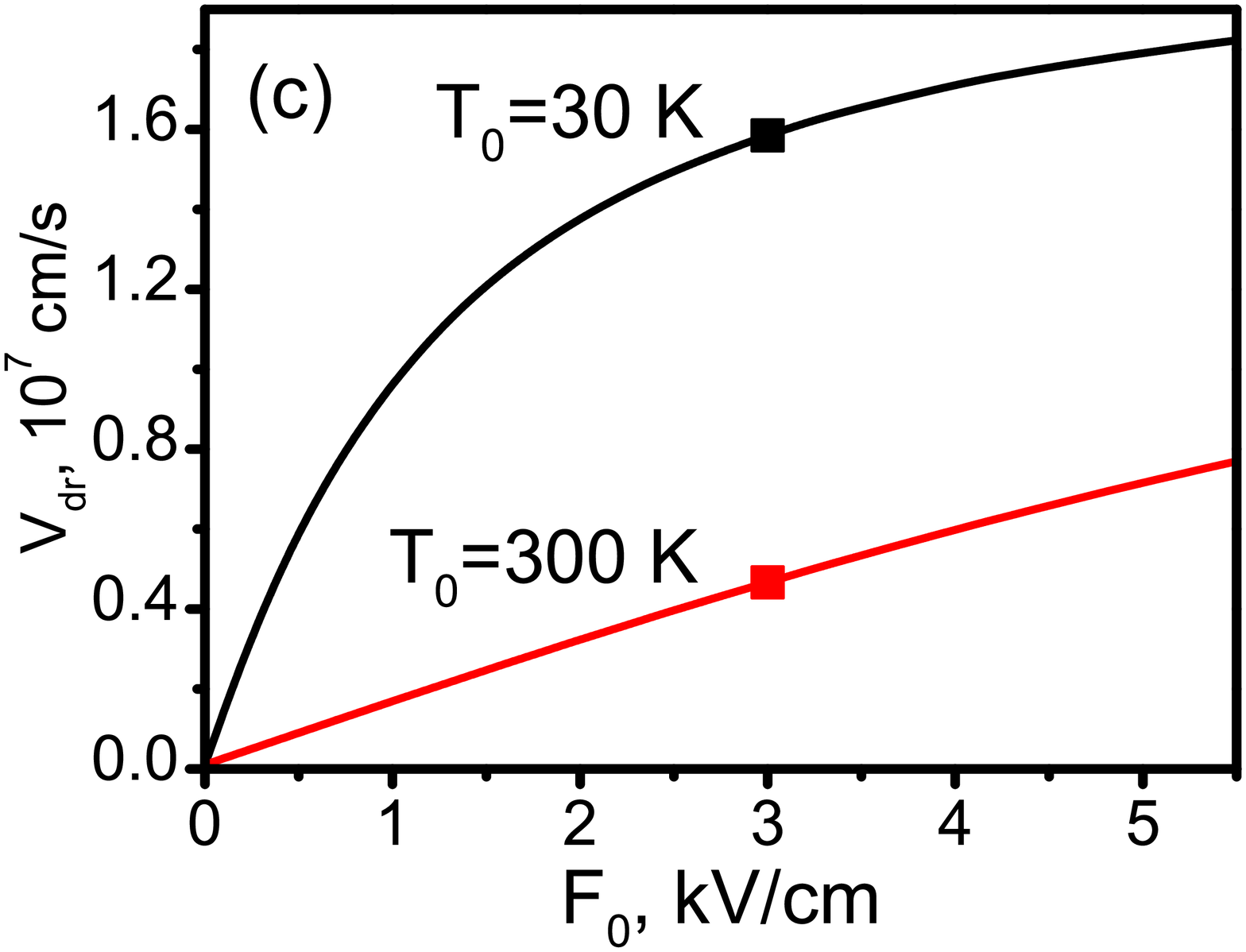}
\caption{Spectrum of the high-frequency conductivity of the drifting electron gas
(solid lines).
(a): $Re[\sigma_{\omega,q}]$. (b): $Im[\sigma_{\omega,q}]$.
The parameters $F_0$, $q$, $T_{0}$  are the same as in Fig.~\ref{fig1}.
In the inset of panel (a): magnified high-frequency region of interest. Dashed-dotted lines are obtained
at $F_{\omega,q}=0.015\times F_{0}$. (c): steady-state dependencies of the drift velocity vs electric field.}
\label{fig2}
\end{figure}
A typical spectral dependence of the HF conductivity in the THz frequency range
for the drifting electron gas is illustrated in Fig.~\ref{fig2} for a given $|q|$.
In this and other figures we show the ratio $\sigma_{\omega,q}/en_{e}$, which is
 the {\it specific} conductivity per one electron~\cite{comment-1}.
Comparing the presented results for the frequency regions $\omega>0$ and $\omega <0$, we
see that the drift of the electrons in the stationary field leads to a strong
non-reciprocal effect in the dynamic conductivity: indeed a change of the sign of $q$ (which is equivalent to a
change of the sign of $\omega$ keeping $q$ unchanged)
strongly modifies the frequency dependence of the HF conductivity.
This corresponds to essentially different responses of the electron gas
to the electric field waves propagating along and against the electron drift.

Another remarkable feature of the frequency dispersion of the HF conductivity of
the drifting electrons is nonmonotonous behavior of both $Re[\sigma_{\omega,q}]$
and $Im[\sigma_{\omega,q}]$ with a set of  "frequency windows", where
the real part of the HF conductivity, $Re[\sigma_{\omega,q}]$, becomes negative.
To illustrate the importance of such frequency windows, we
consider the density of the electric power received by the electrons from the
alternating field, ${\cal P} = {\tilde j}(z,t) \times {\tilde F}(z,t)$.
Using Eqs.~(\ref{j-s}) from Appendix we obtain for the time- and space-averaged
power: $\langle {\cal P} \rangle = \frac{1}{2}\overline{\sigma}_{\omega,q} F_{\omega,q}^2 \cos(\Delta \varphi_{\omega,q})
= \frac{1}{2} Re[\sigma_{\omega,q}] F_{\omega,q}^2$.
As mentioned above, the dissipative electron motion generates an alternating current
with a phase shift, $\Delta \varphi_{\omega,q}$, with respect to the external alternating field.
This phase shift is responsible for attenuation/amplification of the external
alternating signal: if $\Delta \varphi_{\omega,q}$ is such that $\cos(\Delta \varphi_{\omega,q}) > 0 $,
the electrons dissipate the electrical power, if $\cos(\Delta \varphi_{\omega,q}) < 0$ (i.e.
$Re[\sigma_{\omega,q}] < 0 $) the electrons supplies the power to the alternating field
at the expense of the stationary field and current: this means that an {\it amplification of the external field} will take place.
In the case of Fig.~\ref{fig2}, for the frequencies $0.2$ and $1$ THz
indicated by the points $A$ and $C$, $\cos(\Delta\varphi_{\omega,q})<0$
and the amplification is obtained, while for the frequency $0.5$ THz (point B),
$\cos(\Delta\varphi_{\omega,q})>0$ and  the field $\tilde F$ is attenuated.

\begin{figure}[h]
\includegraphics[width=0.47\textwidth]{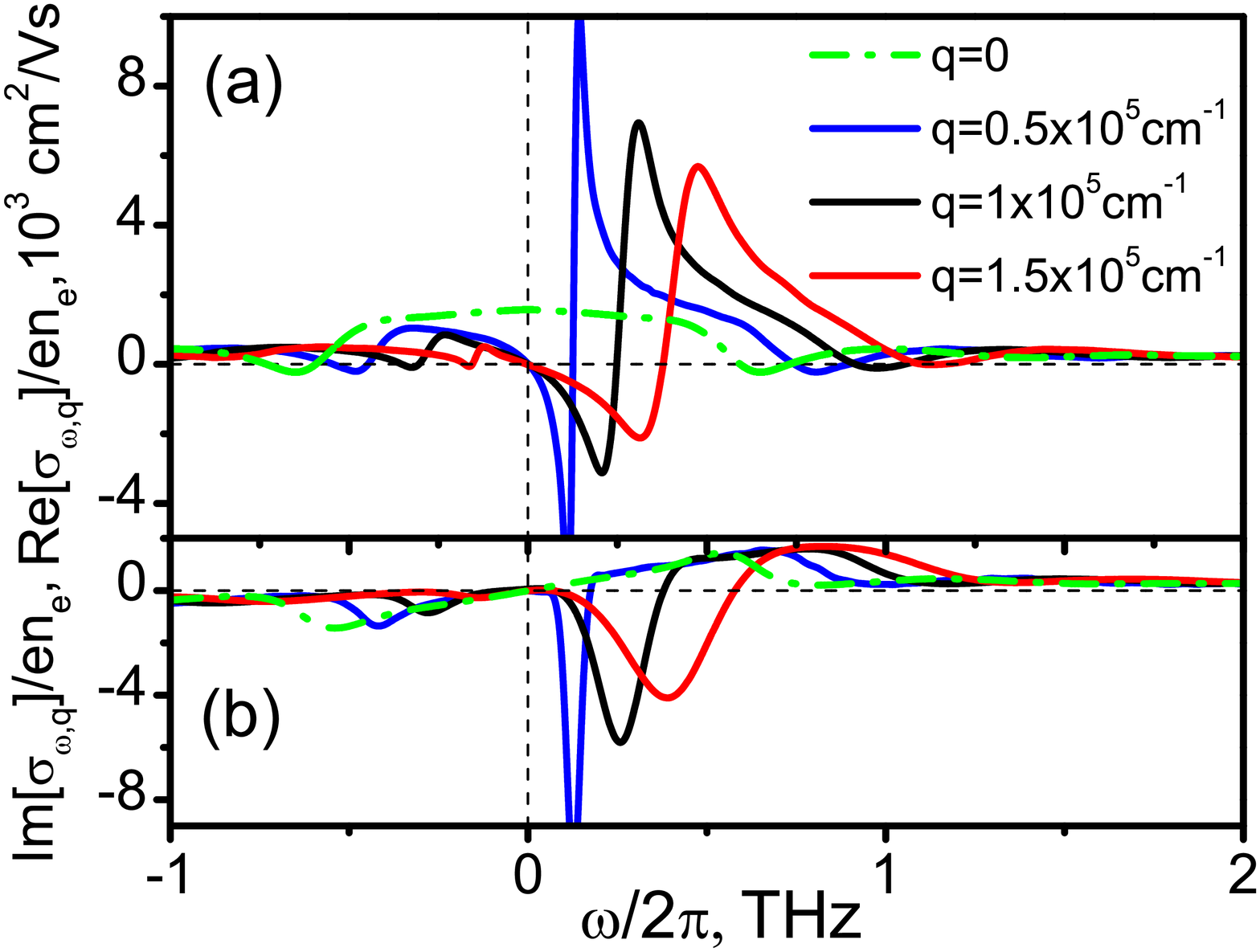}
\vskip 0.1 cm
\includegraphics[width=0.23\textwidth]{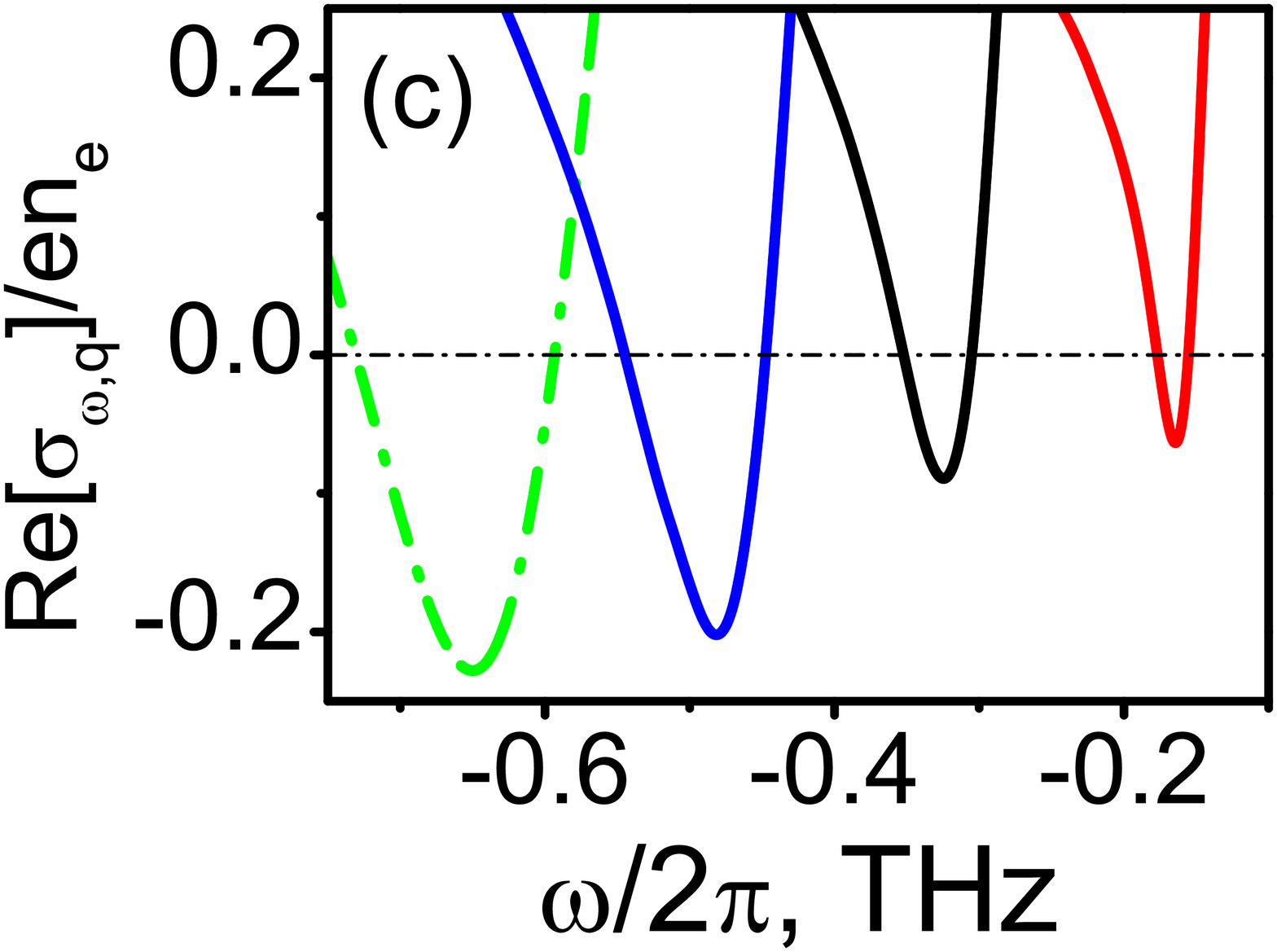}
\includegraphics[width=0.23\textwidth]{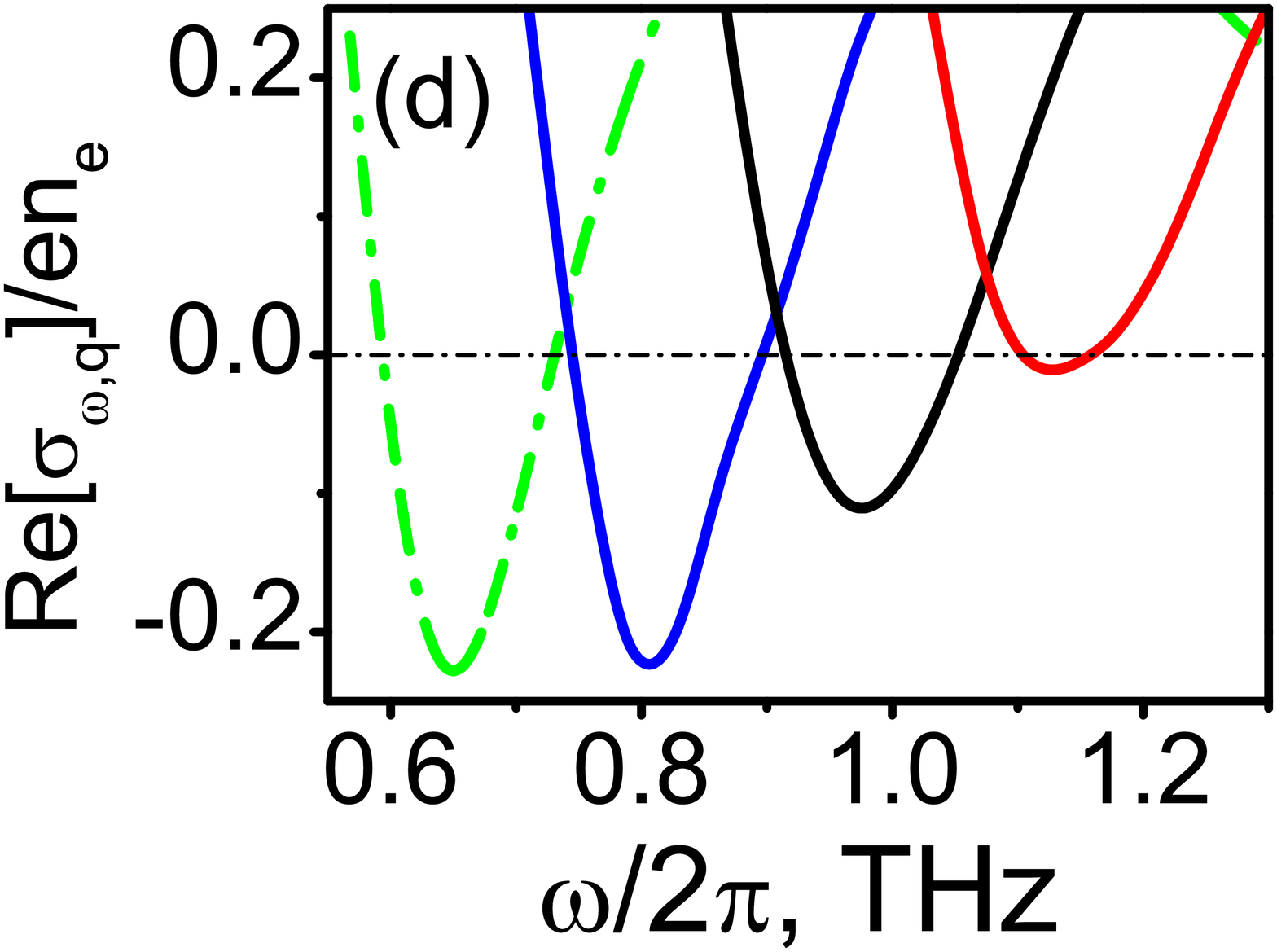}
\caption{Spectra of the high-frequency conductivity of the drifting electron gas
at three different wave vectors $q=0.5,\,1,\,1.5 \times 10^5$ cm$^{-1}$ (curves 1, 2, 3,
respectively). (a): $Re[\sigma_{\omega,q}]$. (b): $Im[\sigma_{\omega,q}]$.
Dash-dotted curves are for $\sigma_{\omega, 0}$.
The panels (c) and (d) show the magnified frequency dependencies $Re[\sigma_{\omega,q}]$
for the windows related to the space-dependent OPTT resonance.
Results are presented for $T_{0}=30$ K and  $F_{0}=3$ kV/cm. }
\label{fig3}
\end{figure}

The physical explanations  of the appearance of the negative HF conductivity, $Re[\sigma_{\omega,q}]$,
are different for the low frequency window and the windows at higher frequencies.
The low-frequency window is characterized by a large effect of
the negative HF conductivity: it can be treated as a manifestation of the well known
Cherenkov effect i.e. an amplification of a wave by electrons drifting with
velocity exceeding the phase velocity of this wave. The Cherenkov amplification
occurs only for waves propagating along the direction of the electron drift.
For example, at
$\omega/2\pi=0.2$ THz and $q=10^{5}$ cm$^{-1}$ corresponding to the point A in
Fig.~\ref{fig2}, the phase velocity, $\omega/q$, is equal to $1.2\times 10^{7}$ cm/s, while
calculations  give a drift velocity $V_{dr}=1.6\times10^{7}$ cm/s at the
stationary field $F_{0}=3$ kV/cm (see Fig.~\ref{fig2}(c)).
The Cherenkov effect in the frequency dependent HF conductivity with
a spatial dispersion is of general character. The dependence of this effect on the
wavevector $q$ and a widening of the corresponding window are illustrated in
Fig.~\ref{fig3}. We remark that the treatment of
this effect can be made even in the framework of the simplified space-dependent
hydrodynamic model. However, this
treatment leads to the divergence of $\sigma_{\omega,q}$ at $\omega=V_{dr} q$.
In contrast, the Monte Carlo method provides a finite results for
$\sigma_{\omega,q}$ and the correct determination of the frequency window
of the Cherenkov effect.

The windows with $Re[\sigma_{\omega,q}] <0$ at higher frequencies
are characterized by a smaller, but noticeable,  effect on the  negative HF conductivity
(see also the panels~\ref{fig3}(c) and ~\ref{fig3}(d)). The physical reason of this effect is the
{\it space-dependent} OPTT resonance, when the electrons oscillate in the nearly-streaming
regime in real and momentum spaces resonantly with
the space- and time-dependent electric field. The space-dependent OPTT resonance
can occur at both signs of $q$, i.e., for the electric wave propagating along as well as
against the electron drift.
At increasing wavevector $q$, these frequency windows are shifted by the
factor $V_{dr} |q|$, as seen from Fig.~\ref{fig3}.
 The amplitudes of the negative HF conductivity under
the space-dependent  OPTT resonance are about one order of magnitude smaller than in the Cherenkov regions.

\begin{figure}[h]
\includegraphics[width=0.47\textwidth]{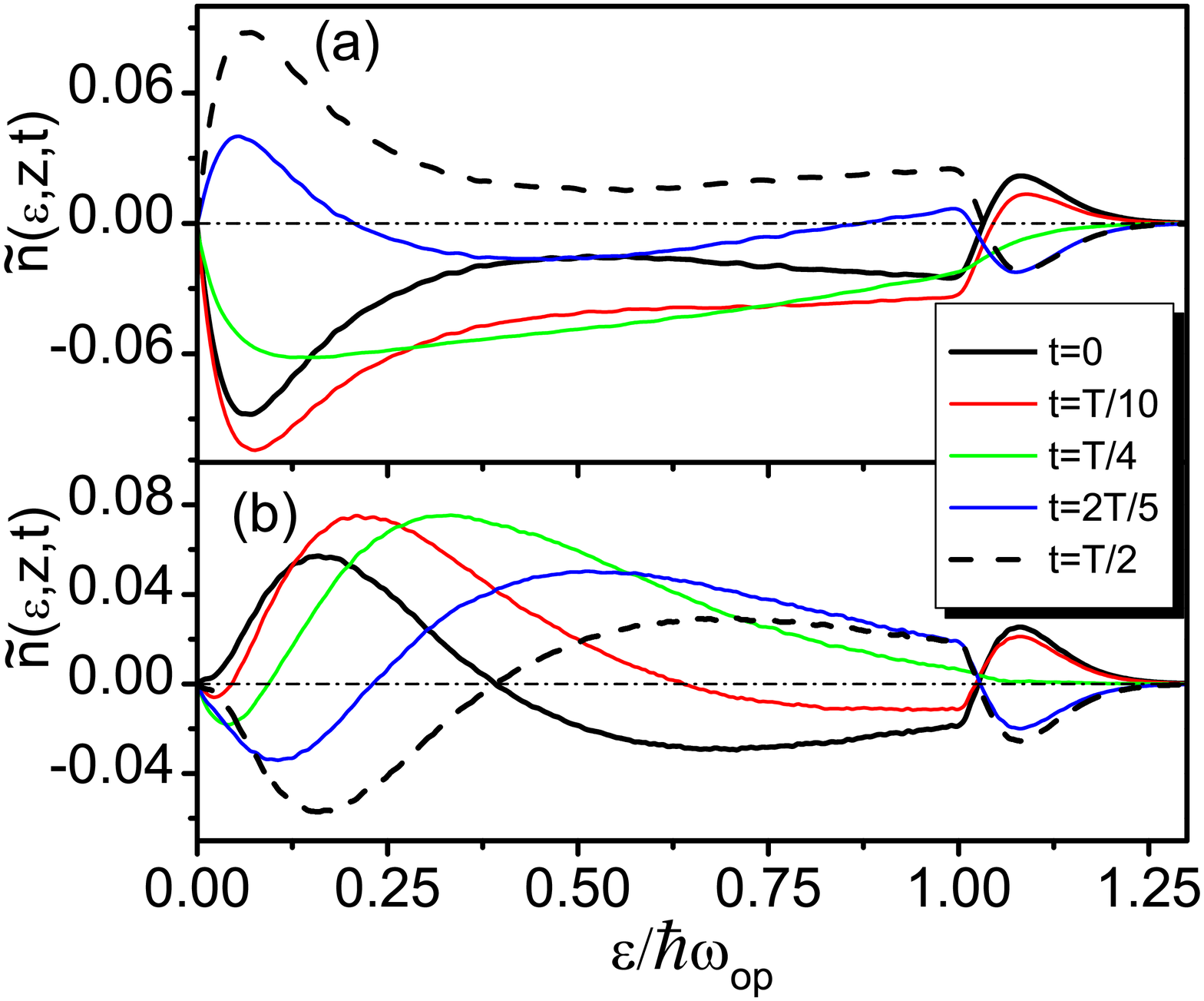}
\includegraphics[width=0.23\textwidth]{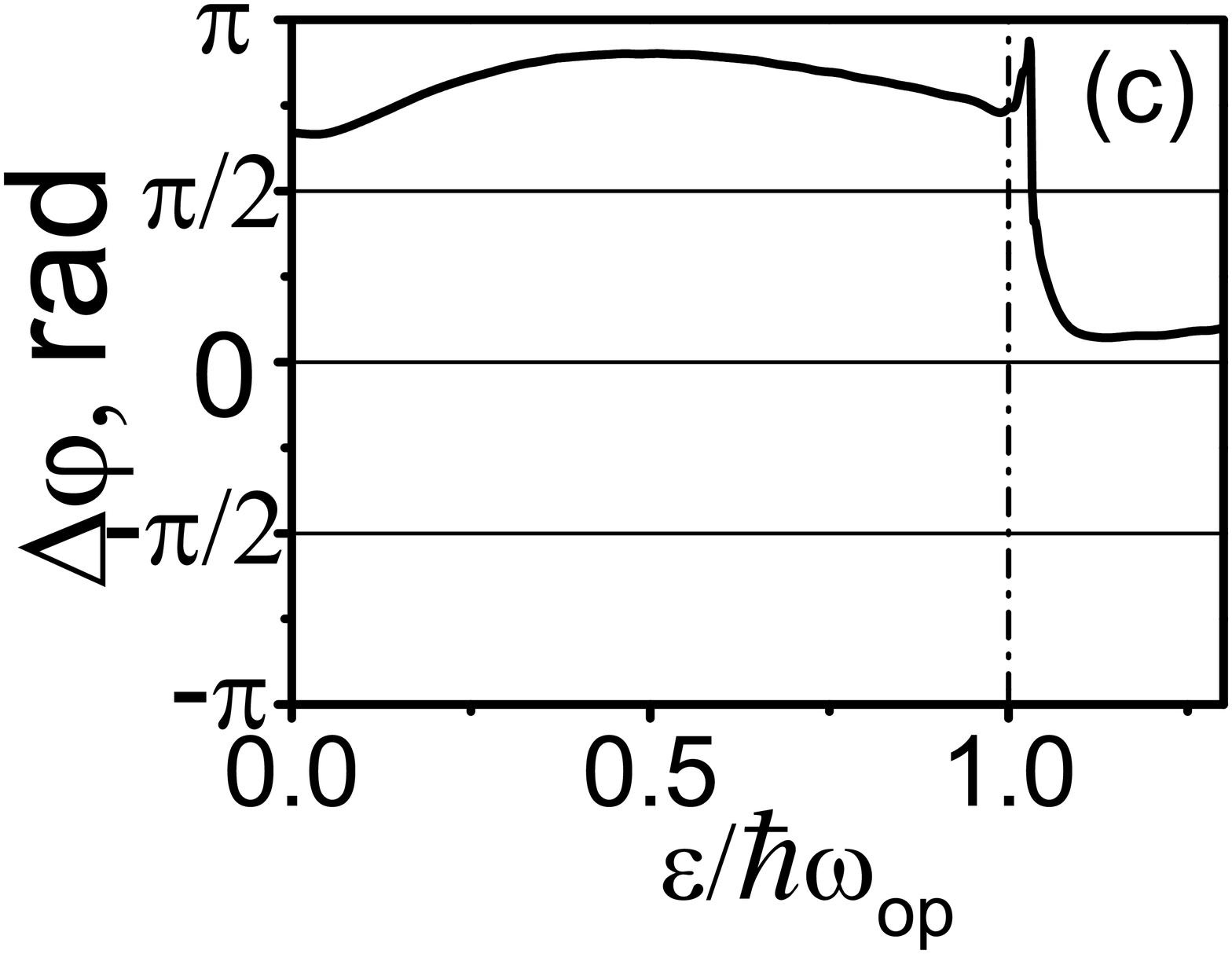}
\includegraphics[width=0.23\textwidth]{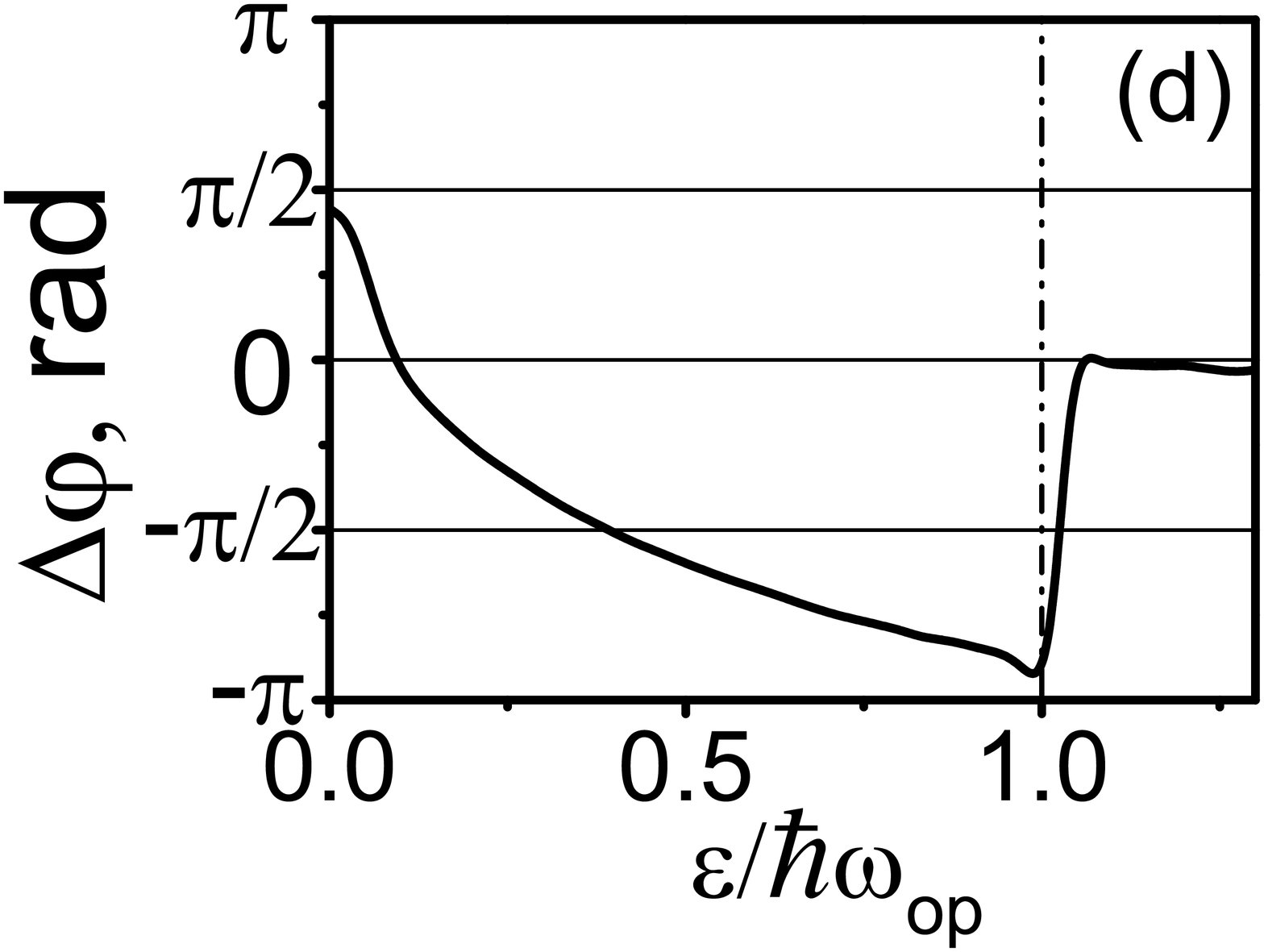}
\caption{Electron density defined by Eq.~(\ref{n-epsilon})  as a function of the
energy,  $\epsilon$,  at a given coordinate, $z=0$, and different moments of time
$t=0, T/10, T/4, 2T/5, T/2$ .
(a): $\omega/2\pi=0.2$ THz; (b): $\omega/2\pi=1$ $THz$. The panels (c) and (d) show the energy dependencies of the phase shift,
$\Delta\varphi(\epsilon)$ for the case of $\omega/2\pi=0.2$ THz and  $\omega/2\pi=1$ THz, respectively. For both panels $q=10^{5}$ cm$^{-1}$. Other parameters are the same as in Fig.~\ref{fig3}.}
\label{fig4}
\end{figure}

As seen from Fig.~\ref{fig3} (a) these resonances vanish with the increasing
of the wavevector $q$. The spectra of $Im[\sigma_{\omega,q}]$
also exhibit a nontrivial behavior (see Fig.~\ref{fig3} (b)).
To understand the differences between the negative HF
conductivity of Cherenkov-type and under the OPTT resonance,
we analyzed the dynamics of the electron gas in both real and momentum spaces.
Such dynamics can be described through the spatial and temporal dependencies of
the average  electrons density with given longitudinal $P_{z}$ and transversal
$P_{\perp}$ momenta with respect to the electric field direction. To illustrate the obtained results, here
we present the density of the electrons with $P_{\perp}(P_x, P_y) =0$ and a given energy
$\epsilon = P_{z}^2/2 m^*$ in the form:
\begin{equation} \label{n-epsilon}
{\tilde n}(\epsilon,z,t) = {\tilde n}_{\omega, q}(\epsilon)
\cos (qz-\omega t+\Delta\varphi_{\omega, q}(\epsilon))\,,
\end{equation}
where ${\tilde n}_{\omega, q}(\epsilon)$ and
$\Delta \varphi_{\omega, q}(\epsilon)$ have been obtained by the Monte Carlo simulations.
In Fig.~\ref{fig4} the dependence of ${\tilde n}(\epsilon, z, t)$ on $\epsilon$
is shown at a given spatial coordinate $z$ for two values of the frequency
corresponding to the windows with the Cherenkov effect (Fig.~\ref{fig4} (a)) and
the OPTT resonance (Fig.~\ref{fig4} (b)).

From Eq.~(\ref{n-epsilon}) and Fig.~\ref{fig4} (a) it follows that
in the Cherenkov frequency window the alternating field of Eq.~(\ref{AF})
induces a {\it traveling wave} of the electron concentration in the real
space and a kind of {\it standing wave} in the energy/momentum space.
In the Fig.~\ref{fig4} (c), the phase shift $\Delta \varphi (\epsilon)$
corresponding to the Cherenkov frequency window is presented: for most of the
electrons having energy $\epsilon < \hbar \omega_{op}$ (the so-called passive region)
this phase shift exceeds $\pi/2$, thus, according to the above analysis,
these electrons amplify the external electric wave. The minority of
electrons with energy $\epsilon > \hbar \omega_{op}$ (the so-called active region) have a phase shifts smaller than $\pi/2$ and contributes to the absorption of the electric wave.

In the frequency windows corresponding to the OPTT resonance (Fig.~\ref{fig4} (d)),
the electric wave of Eq.~(\ref{AF}) is accompanied by oscillations of the electron distribution
in the form of {\it traveling} waves in both the real and
the energy/momentum spaces. The Fig.~\ref{fig4} (d) shows that
the phase shift of these oscillations varies from $\pi/2$ to $-\pi$ depending
on the electron energy. As a consequence, only high-energy electrons in the passive
region $\epsilon < \hbar \omega_{op}$ amplify the external wave. The temporal dynamics of the electron distribution
in the active region ($\epsilon > \hbar \omega_{op}$) is similar to that
of the Cherenkov frequency window. These results  qualitatively explain the distinction
of the effects of the negative HF conductivity  for
the Cherenkov and OPTT resonance frequency windows.

A similar behavior of  the spectra of the HF conductivity
with the wavevector dependence was obtained using an approximate solution of the Boltzmann transport equation
for two-dimensional electron gas in a polar material\cite{Korotyeyev}.

\section{Discussion}

\begin{figure}[h]
\includegraphics[width=0.45\textwidth]{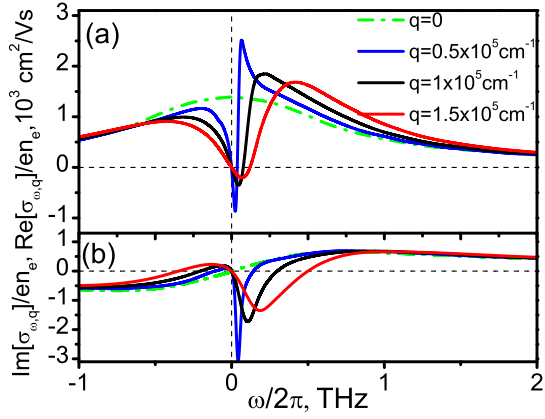}
\caption{The same as in Fig.~\ref{fig3} for $T_{0}=300\, K$.  }
\label{fig5}
\end{figure}

For observation of the negative HF conductivity effects  of
the Cherenkov and OPTT resonance types, low lattice temperatures are
favorable. The Cherenkov effect is less sensitive to the temperature
and exists even at $300$~K as illustrated in Fig.~\ref{fig5}, though it
is realized in a narrower frequency region, because of a smaller drift velocity,
$V_{dr} \approx 0.5\times 10^{7}$ cm/s (see Fig.~\ref{fig2}(c)) at $F_{0}=3$ kV/cm. It is clear that at room temperature this effect also is less pronounced than at
low temperatures (compare with Figs.~\ref{fig3}). The OPTT resonances are present only  at low temperatures,  typically lower than {\it nitrogen} temperature,
i.e. $T_{0} \leq 77$ K.

\begin{figure}[t]
\includegraphics[width=0.45\textwidth]{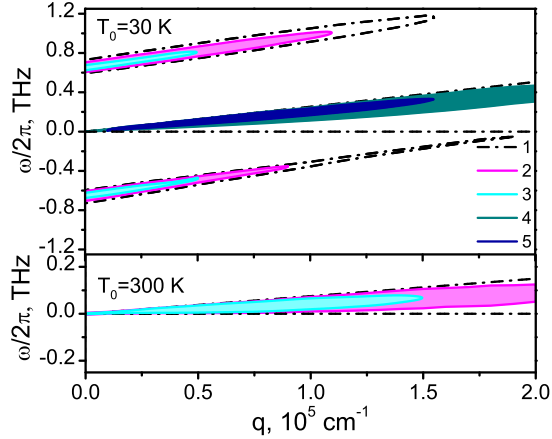}
\caption{Contour plots of $Re[\sigma_{\omega,q}]/en_e$ in the $\{\omega, q \}$-plane.
Curves 1, 2, 3, 4 and 5 restrict the regions of $\{\omega,q\}$,
where $Re[\sigma_{\omega,q}]/en_e <0,\, -100,\, -200,\, -1000,\, -2000$ cm$^{2}$/Vs,
respectively. $F_{0}=3$ kV/cm. }
\label{fig6}
\end{figure}

For the given parameters of the GaN crystal and temperature,
the HF conductivity is dependent on two quantities: $\omega$ and $q$.
Therefore, to characterize the negative HF conductivity effect and
possible amplification of an external wave, one can use
the $\{\omega,q\}$-plane and plot the set of {\it isolines} corresponding
to certain values of $Re[\sigma_{\omega,q}]$. Such a mapping is
presented in Fig.~\ref{fig6} for
$Re[\sigma_{\omega,q}] \leq 0$ at $T_{0}=30$~K and $300$~K.
In particular, the isolines corresponding to $Re[\sigma_{\omega,q}]=0$
separate the $\{\omega, q\}$-regions with negative HF conductivity.
For the case of the Cherenkov effect, this region is the unlimited sector between
the lines $\omega=0$ and $\omega= V_{dr}q$ at $q>0$. For $T_{0}=30$~K and  wavevectors
$q \leq 2\times10^{5}$ cm$^{-1}$, the negative HF conductivity
occurs in the frequency range $0 \div 0.45$ THz. In this frequency
range the {\it specific negative HF conductivity} can reach values of
several thousands of cm$^{2}$/Vs.
For $T_{0}=300$ K, the negative HF conductivity values
are of the order of several hundreds of cm$^{2}$/Vs at frequencies lower than
$0.15$~THz.  Such a suppression of the Cherenkov effect is due to the decrease
of the drift velocity at room temperature.

From Fig.~\ref{fig6}, one can see that at $T_{0}=30$~K the spatially dependent
OPTT resonance  and the negative  HF conductivity occur in a wide frequency range from
0.6 to 1.2~THz, when the wavevector varies from $0$ to $1.5\times 10^{6}$~cm$^{-1}$.
At $q=0$, i.e. in the absence of space dependence of the alternating
field of Eq.(\ref{AF}), the negative HF conductivity due to the OPTT resonance is
possible only in the narrow frequency range of $0.6\div 0.73$~THz.
However, the maximum effect is realized at $q=0$ and $\omega/2\pi=0.65$~THz where
$Re[\sigma_{\omega,q}/en_e]=-230$~cm$^{2}$/Vs.

We suggest that the discovered features associated
with the response of the drifting electrons to {\it high-frequency and spatially
nonuniform electromagnetic fields} are of general character.
Indeed, similar features of the electron response were found for different
drifting electron systems: two-dimensional electrons~\cite{Kempa, Mikhailov},
electrons in graphene strips~\cite{Mikhailov-Gr}  and  two-dimensional electrons in GaN quantum wells~\cite{Korotyeyev}.
Very recently,~\cite{HF-sigma-2018} an oscillation behavior of
the HF conductivity and frequency windows with its negative values
were found and explained in the collisionless limit for structure with GaAs quantum wells.
The important condition to obtain these effects is an anisotropy of the distribution
function of the drifting electrons.
For example, in GaAs quantum wells,  electric fields of order of a few kV/cm (0.5 - 2 kV/cm) induce enough  anisotropic distribution of electrons  to provide a negative $Re[\sigma (\omega, q)]$ in the THz frequency range at $q$ of the order of
$10^5$ cm$^{-1}$ (see Ref. \cite{HF-sigma-2018}).

The dependence of $\sigma_{\omega,q}$ on the wavevector $q$, i.e. the
spatial dispersion, becomes particularly important for samples with submicron- and nanosized structuring. Indeed,
a plane electromagnetic wave illuminating a  nonuniform sample
induces electric field components varying both in space and time,
which interact with the electrons. The spatial dependence of these field components
is defined by the characteristic scales of the structuring of the sample.
Examples of such nonuniform structures are grating-gated semiconductor
structures, surface-relief grating, plasmonic and metamaterial nanodevices,
etc. (see review in Ref.~[\onlinecite{Plasmonics}]). These structures
can be used for different applications, including detecting
and emitting devices of far-infrared and terahertz
radiation.

The knowledge of $\sigma_{\omega,q}$ is also important for the electrodynamic
modeling of the high-frequency characteristics such as transmission, reflection
and absorption. Moreover, the spatially
dependent high-frequency conductivity is directly related
to the plasmonic properties of the electron gas.

In conclusion, using Monte Carlo simulations of the electron motion in
stationary and space- and time-dependent electric fields, we
have determined the wavevector dependence of the HF conductivity, $\sigma_{\omega,q}$,
for compensated GaN samples. In particular, we have found that
the spatially dependent HF conductivity of the drifting electrons can be negative
under stationary electric field of moderate amplitude ($2..5$ kV/cm).
This effect is realized in some frequency windows.
The physics underlying this negative HF conductivity
is different for the low-frequency and the high-frequency windows.
The low-frequency windows are due to the
Cherenkov mechanism. The detailed analysis has shown that
the alternating field induces a {\it traveling wave} of the electron concentration
in the real space and a kind of {\it standing wave} in the energy/momentum space.
The high-frequency windows  are explained by
the OPTT resonances. For this case the alternating field is accompanied by
oscillations of the electron distribution in the form of {\it traveling}
waves in both the real and the energy/momentum spaces. For the observation of
both types of negative HF conductivity effects, low lattice temperatures are favorable. Finally, the negative HF conductivity can be used
to amplify electromagnetic waves at the expense of the energy of the stationary
field and current.

\section{Acknowledgement}
This work is supported by the  Ministry of Education and Science of Ukraine (Project M/24-2018) and German Federal Ministry of
Education and Research (BMBF Project 01DK17028).

\section{Appendix}

An electron subjected to the electric field $F_0+\tilde{F}(z,t)$
and undergoing scatterings by defects and phonons moves along
a complex trajectory in the three-dimensional real space.
To find the alternating electric current induced by the
field $\tilde{F}(z,t)$, we analyze the projection of
the electron trajectory along the $OZ$-axis, i.e. the dependence
$z_{e}(t)$ with $z$ being the electron coordinate and $t \geq 0$, i.e. the projection $z_{e}(t)$ lies
in the right-half of the $\{t,z\}$-plane.
We discretize this half-plane with rectangular cells of height $\Lambda=2 \pi/q$,
and  width $T=2 \pi/\omega$, where $\Lambda$ and $T$ are
the spatial and temporal periods of the field given by Eq.~(\ref{AF}).
A generic $\{{\cal V}_{t}, {\cal V}_{z}\}$-cell is defined as
$$T{\cal V}_{t} <t_{{\cal V}_t}<T ({\cal V}_{t}+1), \, \Lambda{\cal V}_{z} <z_{{\cal V}_z}<\Lambda({\cal V}_{z}+1)$$
with $ {\cal V}_{t}= 0,1,2...N_T-1$ and ${\cal V}_{z}=0,\pm 1,\pm 2...$. Here $N_T$ indicates the number of simulated time periods.
Then we divide each cell  into small meshes
of sizes $T/M_T\times \Lambda/M_Z$, so that the $\{\nu_t, {\cal V}_t; \nu_z, {\cal V}_z\}$-mesh
is determined as:
$$T({\cal V}_t+\nu_t/M_T)<t_{{\cal V}_t,\nu_t}<T({\cal V}_t+(\nu_t+1)/M_T)$$
$$\Lambda({\cal V}_{z}+\nu_z/M_Z)<z_{{\cal V}_{z},\nu_z}<\Lambda ({\cal V}_{z}+(\nu_z+1)/M_Z)$$
with $\nu_t =0,1,2,...M_T-1$ and  $\nu_z=0,1,2,...M_Z-1$; the numbers $M_Z,\,M_T$
are integer and large.
Note that the temporal and spatial periodicities of the external signal imply that all electron characteristics
should have the same periodicity. This means that any mesh corresponding to the same
spatial phase, $qz$, and temporal phase, $\omega t$,  are
equivalent  within a factor $2 \pi \times\, integer$

The calculations of the electron current requires the simulation of electron trajectory also in the {\it momentum} space.
Here, we restrict ourself to the electron current component in the direction of the applied electric field thus we discretize the momentum space along the OZ-direction as follows:
\begin{eqnarray}
-P_{max}(1-(\nu_p+M_{P})/M_{P})<P_{z,\nu_{p}}<-P_{max}\times\nonumber\\
(1-(\nu_p+1+M_{P})/M_{P})\nonumber
\end{eqnarray}
where  $\nu_p =-M_{P}...M_{P}-1$ and $P_{max}$  is selected so that the probability of finding an electron with momentum $P_{z}=P_{max}$ and higher is negligible.

To follow the periodic electron motion induced by the alternating field during the simulation of a sufficiently long trajectory in the $\{z, P_{z}\}$-phase space, we count in each $\{\nu_t, \nu_z\}$-mesh the number of electron appearances ${\cal N}_{\nu_{p}, \nu_t, \nu_z}$ in all meshes with equivalent spatial, $qz_{\nu_{z}}$, and temporal, $\omega t_{\nu_{t}}$, phases and recorded the corresponding momentum projection, $P_{z,\nu_{p}, \nu_t, \nu_z}$. Having these data we can calculate the spatial- and temporal- dependence of the current density, $j_{z}(z_{\nu_{z}},t_{\nu_{t}})$,  within one spatial and temporal period as:
\begin{equation}
\label{Current}
J_{z}(z_{\nu_{z}},t_{\nu_{t}})=-\frac{e}{m^{*}}n_{e}M_{T}M_{z}\sum_{\nu_{p}}P_{z,\nu_{p}, \nu_t, \nu_z}W_{\nu_{p}, \nu_t, \nu_z},
\end{equation}
where $W_{\nu_{p}, \nu_t, \nu_z}$ is the probability to find electron in the $\{\nu_t, \nu_z\}$-mesh with momentum projection, $P_{z,\nu_{p}}$ that is  $W_{\nu_{p}, \nu_t, \nu_z}={\cal N}_{\nu_{p}, \nu_t, \nu_z}/\sum_{\nu_{p}, \nu_t, \nu_z} {\cal N}_{\nu_{p}, \nu_t, \nu_z}$.

By averaging Eq.~(\ref{Current}) with respect to the coordinate and time we can obtain the steady-state current density, $J_{z, 0}(F_{0})$,
\begin{equation}
\label{Current_dc}
J_{z,0}=-\frac{e}{m^{*}}n_{e}\sum_{\nu_{p},\nu_t, \nu_z}P_{z,\nu_{p}, \nu_t, \nu_z}W_{\nu_{p}, \nu_t, \nu_z},
\end{equation}
as well as different Fourier harmonics of the alternating current, $\tilde{j}(z,t)=J_{z}(z,t)-J_{z,0}$.
For example, the first-order Fourier harmonic describing the linear response can be calculated as:
\begin{eqnarray}
\label{Current_ac1}
\left[
\begin{array}{c}
Re[j_{\omega,q}] \\
Im[j_{\omega,q}]
\end{array}
\right]
=-\frac{2e}{m^{*}}n_{e}\!\!\sum_{\nu_{p},\nu_t, \nu_z}P_{z,\nu_{p}, \nu_t, \nu_z}W_{\nu_{p}, \nu_t, \nu_z}
\times\! \nonumber\\
\left[
\begin{array}{c}
\cos(qz_{\nu_{z}}-\omega t_{\nu_{t}}) \\
\sin(qz_{\nu_{z}}-\omega t_{\nu_{t}})
\end{array}
\right]
\end{eqnarray}
The calculated values of $J_{z,0}$, $Re[j_{\omega,q}]$ and $Im[j_{\omega,q}]$ parametrically depends on the magnitude of the stationary field $F_{0}$. In the small-signal limit, $F_{\omega,q}/F_{0}<<1$, the ratios of  $Re[j_{\omega,q}]/F_{\omega,q}$ and $-Im[j_{\omega,q}]/F_{\omega,q}$ become independent of the alternating field amplitude, $F_{\omega,q}$ and give the real, $Re[\sigma_{\omega,q}]$, and imaginary, $Im[\sigma_{\omega,q}]$, parts of the complex HF conductivity, respectively.
In this case the alternating current $\tilde{j}(z,t)$ has the almost harmonic form:
\begin{equation} \label{j-s}
\tilde{j}(z,t) \approx {\overline \sigma}_{\omega,q} F_{\omega,q}
\cos (qz-\omega t + \Delta \varphi_{\omega,q})\,,
\end{equation}
 where  ${\overline \sigma}_{\omega,q}=\sqrt{Re[\sigma_{\omega,q}]^2+Im[\sigma_{\omega,q}]^2}$
 and the phase shift between the alternating field and the current is $\Delta \varphi_{\omega,q}=\arctan(Im[\sigma_{\omega,q}]/Re[\sigma_{\omega,q}])$.

To prove the accuracy of the simulation,
we checked the convergence of the calculations varying the numbers and sizes of
the cells and the meshes used in the simulation of space- and time dependent
electron transport.
\begin{figure}[h]
\includegraphics[width=0.45\textwidth]{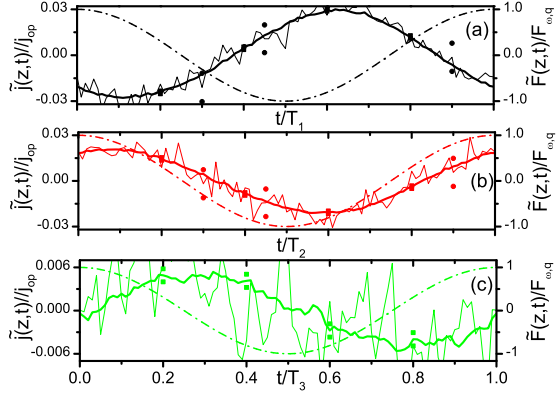}
\caption{Time dependencies of the alternating current
calculated for $N_{T}=10^{6}$ (thin curve) and $N_{T}=10^{8}$ (thick curve).
(a), (b) and (c) correspond to the frequencies  $\omega_A,\,\omega_B$ and $\omega_C$ presented
in the text. The time is normalized to the corresponding period of the alternating
field whose time-dependence is shown by the dash-dotted lines.
Results are presented for $z=0$ and $F_{\omega,q}=0.1 \times F_{0}$. }
\label{fig7}
\end{figure}
As an example, Fig.~\ref{fig7} presents the alternating currents obtained using different numbers of the temporal period, $N_T$, for three frequencies (marked by points in Fig.\ref{fig2}) of the field $\tilde{F}$.  For the used mesh sizes, $\Lambda/M_Z,\,T/M_T$ we set
$M_Z=M_T=100$. As seen from Fig.~\ref{fig7}, the two curves calculated for
$N_{T}=10^{6}$ and $N_{T}=10^{8}$ coincide within a small statistical error
for two cases of low frequencies $\omega_A/2\pi = 0.2$ and
$\omega_B/2\pi = 0.5$~THz (Fig.~\ref{fig7} (a) and (b)).
At higher frequency, $\omega_C/2\pi=1$~THz, only the use of $N_{T}=10^{8}$ is
sufficient to obtain a satisfactory accuracy of the computations (Fig.~\ref{fig7} (c)).

By varying the field amplitude $F_{\omega,q}$ from $0.015\times\,F_0$ to $0.1\times\,F_0$,
we found that the ratio $j_{\omega,q}/F_{\omega,q}$ does not depend
on the amplitude $F_{\omega,q}$ for $F_{\omega,q} \leq 0.1 \,F_0$.

In the previous sections, we have presented results for the $N_{T}=10^{8}$, $M_{T}=100$, $M_{Z}=100$ and
$F_{\omega,q}/F_{0}=0.1$. For such parameters the relative error of the calculation
of $\sigma_{\omega, q}$ does not exceed $1\%$ in the range of investigated frequencies
and wavevectors. The computational parameters related to the {\it momentum} space were
$M_{P}=500$ and $P_{max}=3\times p_{op}$.



\begin{references}
\bibitem{comment-1} Though the dimension of $\sigma_{\omega,q}/en_e$
is $cm^2/V\,s$, this ratio is not an electron mobility.
\end{references}
\end{document}